\documentclass{Lisbon}
\newcommand{\too}{\mathop{\to}\limits_{N_C\to\infty}}
\newcommand{\vpint}{\int\makebox[0mm][r]{\bf --\hspace*{0.13cm}}}
\newcommand{\ds}{\displaystyle}
\newcommand{\vph}{\varphi}
\title{\bf Hamiltonian approach to $\rm QCD_2$ in the axial gauge.}
\author{A.V.Nefediev}
\begin{document}
\maketitle

\abstracts{
The Hamiltonian approach is developed for $\rm QCD_2$ in the limit of infinite 
number of colours $N_C$ ('t Hooft model). Bosonization of the theory is performed 
explicitly and the generalized Bogoliubov transformation for the 
composite boson operators is introduced and used to bring the
Hamiltonian into diagonal form in the two-body sector. The resulting 
theory is re-formulated in terms of effective degrees of freedom 
and describes the free mesons with creation and annihilation operators commuting
in the canonical way. Corrections in $N_C$ to the leading term in the Hamiltonian 
describe the interaction between mesons and can be used to consider their decays.
Chiral properties of the theory are discussed and it is shown that the backward
motion of the chiral pion (Goldstone mode) is not suppressed and should contribute to
the decays amplitudes.}

In 1974 a model for QCD in two dimensions in the limit of infinite number of
colours was suggested by 't~Hooft \cite{'tHooft}. It is described by
the Lagrangian
\be
L(x)= -\frac{1}{4}F^a_{\mu\nu}(x)F^a_{\mu\nu}(x)+\bar q(x)(i\hat{D}-m)q(x),
\label{1}
\ee
where $\hat{D} = (\partial_{\mu} - igA^a_{\mu}t^a)\gamma_{\mu}$, and
the convention for $\gamma$-matrices is $\gamma_0 = \sigma_3$,
$\gamma_1 = i\sigma_2$, $\gamma_5 = \gamma_0\gamma_1$. The large $N_C$
limit implies that $g^2N_C$ remains finite.

For 25 years this model has been under 
intent attention of many theorists since it possesses many features similar to those 
of four dimensional QCD and most of them can be studied analytically. This model 
is a brilliant test-bed for investigating such nonperturbative phenomena as 
confinement and chiral symmetry breaking.

The fact that two-dimensional gluon contains no propagating degrees of freedom
allows to construct a self-consistent Hamiltonian approach.
According to the standard rules one finds the Hamiltonian of the model in the axial
gauge $A_1=0$ to be
$$
H=\int dxq^{+}(x)\left(-i\gamma_5\frac{\partial}{\partial x}+m\gamma_0\right) q(x)
\hspace*{2cm}
$$
\be
\hspace*{2cm}-\frac{g^2}{2}\int dxdy\;q^{+}(x)t^aq(x)q^{+}(y)t^aq(y)\frac{\left|x-y\right|}{2},
\label{2}
\ee
where "dressed" quark field defined by means of the standard Boguliubov-Valatin 
angle $\theta$ is introduced:
\be
q_i(x_0,x)=\int\frac{dk}{2\pi}\left[u(k)b_i(x_0,k)+v(-k)d_i(x_0,-k)\right]e^{ikx}
\label{3}
\ee
$$
u(k)=T(k)\left(1 \atop 0 \right)\quad v(-k)=T(k)\left(0 \atop 1 \right)\quad 
T(k)=e^{-\frac{1}{2}\theta(k)\gamma_1}.
$$

First we shall recall the diagonalization of the model in the 
one-body sector to find the exact form of the "dressed" quarks and to define
the "true" quark vacuum of the model which will appear chirally non-symmetric. 

After arranging the normal ordering in Hamiltonian (\ref{2}) in the given 
basis (\ref{3}) one has the Hamiltonian consisting of three parts: vacuum energy, 
the term quadratic in quark operators $(:H_2:)$ and the one of the fourth power
$(:H_4:)$.

The standard requirement that $H_2$ be diagonal in terms of quark 
creation and annihilation operators leads to the gap equation\cite{Bars&Green}
\be
p\cos\theta(p)-m\sin\theta(p)=\frac{\gamma}{2}\vpint\frac{dk}{(p-k)^2}\sin[\theta(p)-\theta(k)],
\label{6}
\ee
or equivalently to the system of two coupled equations where the "dressed" quark 
dispersive law $E(p)$ is introduced\cite{Bars&Green}: 
\be
\left\{
\begin{array}{c}
E(p)\cos\theta(p)=m+\frac{\ds \gamma}{\ds 2}\ds\vpint\frac{dk}{(p-k)^2}
\cos\theta(k)\\
{}\\
E(p)\sin\theta(p)=p+\frac{\ds \gamma}{\ds 2}\ds\vpint\frac{dk}{(p-k)^2}
\sin\theta(k),
\end{array}
\right.
\label{7}
\ee

It was shown numerically\cite{Ming Li} that even in the chiral limit 
gap equation (\ref{6}) has a
nontrivial solution for $\theta$ which gives stable chirally noninvariant vacuum
state. We shall return to this issue later when discussing the
chiral properties of the model.

It is easy to check that contribution of the $H_4$ term on the "dressed" quarks
states is suppressed by powers of $N_C$. Thus the total Hamiltonian is completely
diagonalized in the one-body sector. Having stopped at this step one can develop a
diagrammatic technique with "dressed" quarks lines and Green's functions 
involved\cite{Bars&Green},
whereas we shall proceed further to perform diagonalization of the model in the mesonic
sector.

From the pioneer work by 't~Hooft we know that the spectrum of the model
consists of mesons, so that it would be very natural to express the
Hamiltonian of the theory in terms of mesonic states.

First of all we introduce the two-body operators 
\be
\begin{array}{c}
B(p,p')=\frac{\ds 1}{\ds\sqrt{N_C}}b_i^{+}(p)b_i(p')\quad 
D(p,p')=\frac{\ds 1}{\ds\sqrt{N_C}}d_i^{+}(-p)d_i(-p')\\
{}\\
M(p,p')=\frac{\ds 1}{\ds\sqrt{N_C}}d_i(-p)b_i(p')\quad
M^{+}(p,p')=\frac{\ds 1}{\ds\sqrt{N_C}}b^{+}_i(p')d^{+}_i(-p)
\end{array}
\label{8}
\ee
and express the Hamiltonian in their terms. 
The commutation relations for the new operators can
be easily found and the only non-vanishing one reads as follows
\be
[M(p,p')M^{+}(q,q')]\too(2\pi)^2\delta(p'-q')\delta(p-q).
\label{9}
\ee

As soon as in the mesonic sector of the theory one deals with the $q\bar q$ pairs
only, so neither isolated quark nor isolated antiquark can be created or annihilated.
Such way operators $B$ and $D$ from (\ref{8}) can not be independent and have to be
related to operators $M$ and $M^+$ somehow. Indeed, it is easy to check that the 
following substitution\cite{Lenz}
\be
\begin{array}{c}
B(p,p')=\frac{\ds 1}{\ds\sqrt{N_C}}\ds\int\frac{\ds dq}{\ds 2\pi}M^{+}(q,p)M(q,p')\\
{}\\
D(p,p')=\frac{\ds 1}{\ds\sqrt{N_C}}\ds\int\frac{\ds dq}{\ds 2\pi}M^{+}(p,q)M(p',q)
\end{array}
\label{10}
\ee
comes through the commutation relations (\ref{9}) so that the Hamiltonian takes the
form
$$
H=LN_C{\cal E}_v+\int\frac{dQdp}{(2\pi)^2}\left[(E(p)+E(Q-p))M^{+}(p-Q,p)M(p-Q,p)\right.
$$
\be
-\frac{\gamma}{2}\int\frac{dk}{(p-k)^2}\left\{2C(p,k,Q)M^{+}(p-Q,p)M(k-Q,k)\right.
\label{11}
\ee
$$
\left.\left.+S(p,k,Q)\left(M(p,p-Q)M(k-Q,k)+M^{+}(p,p-Q)M^{+}(k-Q,k)\right)\right\}\right],
$$
where
\be
\begin{array}{c}
C(p,k,Q)=\ds\cos\frac{\ds\theta(p)-\theta(k)}{\ds 2}\cos\frac{\ds
\theta(Q-p)-\theta(Q-k)}{\ds2}\nonumber\\
\vphantom{.}\\
S(p,k,Q)=\ds\sin\frac{\ds\theta(p)-\theta(k)}{\ds
2}\sin\frac{\ds\theta(Q-p)-\theta(Q-k)}{\ds2}\nonumber.
\label{12}
\end{array}
\ee

Note that, being of the same order in powers of $N_C$, $H_2$ and $H_4$ parts are equally 
important in the two-body sector.

As a nest step let us define mesonic creation and annihilation operators in
the form\cite{we}
\be
\begin{array}{c}
m^{+}_n(Q)=\ds\int\frac{\ds dq}{\ds 2\pi}\left\{M^+(q-Q,q)\vph_+^n(q,Q)+
M(q,q-Q)\vph_-^n(q,Q)\right\}\\
\\
m_n(Q)=\ds\int\frac{\ds dq}{\ds 2\pi}\left\{M(q-Q,q)\vph_+^n(q,Q)+
M^+(q,q-Q)\vph_-^n(q,Q)\right\},
\label{13}
\end{array}
\ee
where the subscript $n$ numerates mesonic states, and $Q$ being the total
momentum of the meson.
The given transformation is nothing
but another Bogoliubov one generalized for the case of composite bosonic operators.
Wave functions $\varphi_+$ and $\varphi_-$ obey the following completeness and
orthogonality conditions
\be
\begin{array}{rcl}
\ds\int\frac{\ds dp}{\ds 2\pi}\left(\vph_+^n(p,Q)\vph_+^{m}(p,Q)-\vph_-^n(p,Q)\vph_-^m(p,Q)
\right)&=&\delta_{nm}\\
&&\\
\ds\int\frac{dp}{2\pi}\left(\vph_+^n(p,Q)\vph_-^{m}(p,Q)-\vph_-^n(p,Q)\vph_+^m(p,Q)
\right)&=&0
\label{14}
\end{array}
\ee
\be
\begin{array}{rcl}
\ds\sum\limits_{n=0}^{\infty}\left(\vph^n_+(p,Q)\vph^n_+(k,Q)-\vph^n_-(p,Q)
\vph^n_-(k,Q)\right)&=&2\pi\delta\left( p-k\right)\nonumber\\
&&\\
\ds\sum\limits_{n=0}^{\infty}\left(\vph^n_+(p,Q)\vph^n_-(k,Q)-\vph^n_-(p,Q)
\vph^n_+(k,Q)\right)&=&0\nonumber,
\label{15}
\end{array}
\ee
that ensures the standard bosonic commutation relation for the operators $m$ and
$m^+$:
\be
\begin{array}{cc}
\left[m_n(Q)m^+_m(Q')\right]=2\pi\delta(Q-Q')\;\delta_{nm}\\
{}\\
\left[m_n(Q)m_m(Q')\right]=\left[m^+_n(Q)m^+_m(Q')\right]=0.
\end{array}
\label{16}
\ee
 
Particular attention should be payed to the sign "minus"
between the two parts of relations (\ref{14}), (\ref{15}), 
which is the direct analogue of the
corresponding sign in the standard bosonic Bogoliubov condition $u^2-v^2=1$.

The Hamiltonian of the models takes the final diagonal form in the given basis 
with ${\cal E}_v$ being the vacuum energy density and $L$ --- volume of the 
one-dimensional $x$-space
\be
H=LN_C{\cal E}_v
+\sum\limits_{n=0}^{+\infty}\int\frac{dQ}{2\pi}Q^0_n(Q)m^+_n(Q)m_n(Q)+
O\left(\frac{1}{\sqrt{N_C}}\right),
\label{17}
\ee
if the wave functions obey the following integral system of equations\cite{Bars&Green}
\be
\left\{
\begin{array}{c}
[E(p)+E(Q-p)-Q_0]\vph_+(p,Q)\hspace*{5cm}\\
\hspace*{1cm}=\gamma\ds\vpint\frac{\ds dk}{\ds (p-k)^2}
\left[C(p,k,Q)\vph_+(k,Q)-S(p,k,Q)\vph_-(k,Q)\right]\\
{}\\

[E(p)+E(Q-p)+Q_0]\vph_-(p,Q)\hspace*{5cm}\\
\hspace*{1cm}=\gamma\ds\vpint\frac{\ds dk}{\ds (p-k)^2}
\left[C(p,k,Q)\vph_-(k,Q)-S(p,k,Q)\vph_+(k,Q)\right].
\end{array}
\right.
\label{18}
\ee

A comment on the physical meaning of the two wave functions 
$\varphi_+$ and $\varphi_-$ is in order: $\varphi_+$ describes the
forward motion in time of the $q\bar q$ pair inside the meson, whereas $\varphi_-$
stands for the backward motion. Numerical simulations show that $\varphi_-$ 
component is suppressed for highly excited mesonic states as well as for rather 
massive quarks\cite{Ming Li}. 

To trace the root of the unusual form of equations (\ref{14}), (\ref{15}) let us 
re-write the system (\ref{18})
in the Shr{\"o}dinger-like matrix form
\be
Q^n_0\left(\vph_{+}^n\atop \vph_{-}^n \right)=
\hat{\cal H}\left(\vph_{+}^n\atop \vph_{-}^n \right)
\ee
and note that matrix Hamiltonian $\hat{\cal H}$ appears non-Hermitian in the
full Hilbert space of arbitrary functions $\varphi_+$ and $\varphi_-$. Nevertheless
this does not lead to a disaster as "physical" wave functions belong to a
restricted space defined by means of projectors
$\Lambda_{\pm} = T(k)\frac{1 \pm \gamma_0}{2}T^+(k)$
and one can check explicitly that all eigenenergies of the system (\ref{18}) are real\cite{we}.
Distorted norm (\ref{14}) is nothing but another reflection of the above space 
restriction.

Let us discuss possible applications of the developed method. We shall concentrate on
the chiral properties of the model.

The general idea of calculation of any matrix element between hadronic states is to
express the corresponding operator in terms of mesonic creation and annihilation
operators $m$ and $m^+$ and to calculate the matrix element explicitly using the standard second
quantization technique.

Chiral condensate is the first example. First of all, note that Bogoliubov 
transformation not 
only changes operators but also re-orders the vacuum state. Luckily the contribution
of this re-ordering has sub-leading order in $N_C$ when a matrix 
element of any bilinear combination of quarks fields is calculated. 
Thus following the way described above one arrives at
\be
\langle\bar{q}q\rangle=
-\frac{N_c}{2\pi}\int\limits_{-\infty}^{+\infty}dk\cos\theta(k).
\label{21}
\ee

Substituting the numerical solution for $\theta$ into
(\ref{21}) one finds nonzero result (as expected for chirally
non-invariant vacuum) which is in agreement with the one found from the sum
rules\cite{Zhitnitsky}. 

Despite of the fact that the bound state equation (\ref{18}) is rather a
subject to numerical study, an appropriate solution for the lowest massless
state can be found analytically in the chiral limit
\be
\vph^{\pi}_{\pm}(p,Q) =\sqrt{\frac{\pi}{2Q}}\left(\cos\frac{\theta(Q-p)-\theta(p)}{2}\pm
\sin\frac{\theta(Q-p)+\theta(p)}{2}\right)
\label{22}
\ee
and this solution is nothing but the chiral pion --- Goldstone mode present in the
spectrum of the model. The two comments are in order here: i) being massless this state 
is not defined in the rest frame (at $Q\to 0$) as clearly seen from equation (\ref{22}); 
ii) for finite $Q$'s both wave functions $\vph^{\pi}_+$ and 
$\vph^{\pi}_-$ are of comparable order of magnitude, so that the $q\bar q$ pair spends
about half time moving forward and about halt time moving backward in time. Thus in contrary to
the higher excited states, for which $\varphi_-$ component is suppressed, in case of pion both wave functions are of the same
importance and none of them can be neglected.

Armed with the $\pi$-meson solution one can find the pionic decay constant 
$f_{\pi}$ defined as
\be
\left.\left\langle\Omega\right.\right|J_{\mu}^5(x)\left|\left.\pi(Q)\right.
\right\rangle=f_{\pi}Q_{\mu}\frac{e^{-iQx}}{\sqrt{2Q_0}}
\label{23}
\ee
to be $\sqrt{N_C/\pi}$. Besides the celebrated 
Gell-Mann-Oakes-Renner relation 
\be
f_{\pi}^2M_{\pi}^2=-2m\langle\bar{q}q\rangle,
\label{24}
\ee
can be readily reconstructed from the system (\ref{18}).

In conclusion we would like to note that the applications of the developed approach 
are not exhausted with the given examples. Mesonic decays and scattering can be
studied if the next-to-leading terms in the Hamiltonian of the theory are taken into
account. This might help, for example, to disclose the role played by pions, as massless Goldstone modes, 
in the hadronic processes.
\smallskip

The author would like to thank the Organizing Committee of the 
17th Autumn School in QCD, where this talk was given, for very warm hospitality.
Financial support of RFFI grants 97-02-16406 and 96-15-96740 and INTAS-RFFI grant IR-97-232
is gratefully acknowledged.


\begin{thebibliography}{99}
\bibitem{'tHooft} G.'t~Hooft, Nucl.Phys. {\bf B75}, 461 (1974)
\bibitem{Bars&Green} I.Bars and M.B.Green, Phys.Rev. {\bf D17},
537 (1978)
\bibitem{Ming Li} Ming Li, Phys.Rev. {\bf D34}, 3888 (1986)
\bibitem{we} Yu.S.Kalashnikova, A.V.Nefediev, A.V.Volodin, Phys.At.Nucl. in press
\bibitem{Zhitnitsky} A.R.Zhitnitsky, Sov.J.Nucl.Phys. {\bf 43},
999 (1986), {\bf 44}, 139 (1986)  
\bibitem{Lenz} F.Lenz and M.Thies, Ann.Phys. (N.Y.) {\bf 208}, 1
(1991)
\bibitem{Birse} Ming Li, L.Wilets and M.C.Birse, J.Phys. {\bf G13},
915 (1987)
\bibitem{Oakes} M.Gell-Mann, R.J.Oakes and B.Renner, Phys.Rev. {\bf 175} (1968) 
2195 
\end{thebibliography}
\end{document}